\documentclass[a4paper]{article}

\usepackage{INTERSPEECH2022}

\usepackage{soul}
\usepackage[dvipsnames]{xcolor}
\usepackage{duckuments}
\usepackage{graphicx}
\usepackage{array}
\usepackage{multirow}
\usepackage{caption}
\usepackage{subfig}
\usepackage{float}

\title{SATTS: Speaker Attractor Text to Speech, \\Learning to Speak by Learning to Separate }
\name{Nabarun Goswami$^1$, Tatsuya Harada$^{1,2}$}
\address{
  $^1$The University of Tokyo, Japan\\
  $^2$RIKEN, Japan}
\email{\{nabarungoswami,harada\}@mi.t.u-tokyo.ac.jp}

\begin{document}

\maketitle
\begin{abstract}

The mapping of text to speech (TTS) is non-deterministic, letters may be pronounced differently based on context, or phonemes can vary depending on various physiological and stylistic factors like gender, age, accent, emotions, etc. Neural speaker embeddings, trained to identify or verify speakers are typically used to represent and transfer such characteristics from reference speech to synthesized speech. Speech separation on the other hand is the challenging task of separating individual speakers from an overlapping mixed signal of various speakers. Speaker attractors are high-dimensional embedding vectors that pull the time-frequency bins of each speaker's speech towards themselves while repelling those belonging to other speakers. In this work, we explore the possibility of using these powerful speaker attractors for zero-shot speaker adaptation in multi-speaker TTS synthesis and propose speaker attractor text to speech (SATTS). Through various experiments, we show that SATTS can synthesize natural speech from text from an unseen target speaker's reference signal which might have less than ideal recording conditions, i.e. reverberations or mixed with other speakers.


\end{abstract}

\section{Introduction}
With the advancement of various deep learning techniques, TTS systems have improved quite a lot. Most modern TTS systems have two parts, a front end synthesizer and a backend vocoder. The synthesizer takes as input text or phonemes and synthesizes an intermediate representation like mel-spectrogram. Examples of such synthesizers are the Tacotron family of synthesizers \cite{wang2017tacotron, shen2018natural, elias2021parallel}, Transformer TTS \cite{li2019transformertts}, FastSpeech \cite{ren2019fastspeech}, etc. The backend vocoders convert the intermediate representations into speech waveforms. Various kinds of vocoders have been proposed including, but not limited to, Wavenet \cite{oord2016wavenet}, WaveGlow \cite{prenger2019waveglow}, MelGAN \cite{kumar2019melgan}, HiFi-GAN \cite{kong2020hifi}, etc. Another family of TTS systems work end-to-end, i.e. convert text directly into waveform without going through an intermediate representation, Methods such as EATS \cite{donahue2021endtoend} and VITS \cite{kim2021conditional} are examples of end-to-end TTS systems. 

These methods have pushed the boundaries of quality of synthesized speech, in fact most single speaker methods have naturalness almost at par with real human speech. However, there is still much to be desired in terms of multi-speaker speech synthesis, especially in zero-shot speaker adaptation setting. Multi speaker setting is usually incorporated into the TTS systems in the form of conditioning speaker embedding. This embedding might be from a look up table in the case when using fixed speaker identities \cite{oord2016wavenet, kim2021conditional}, trained along with TTS model \cite{gibiansky2017deep, wang2018style} or in the form of embeddings extracted from a speaker discriminative model such as speaker verification as in SV2TTS \cite{jia2018transfer} and YourTTS \cite{Casanova2021yourtts}.
The third method of extracting embeddings from a speaker discriminative model allows for zero shot speaker adaptation without changing the parameters of the TTS system at inference time. While such speaker discriminative embeddings have been shown to work well for multi speaker TTS systems, these embeddings are mainly trained to just capture the broad global attributes of different speakers and being agnostic to the specific features of a particular reference speech sample. 

To alleviate this problem, we propose to use speaker attractors \cite{jiang2020speaker} as embeddings for zero shot speaker adaptation. Speaker attractors are high dimensional embeddings which are used for pulling the time-frequency embeddings of a speaker closer to itself in the task of speech separation from a mixture of different speakers. The speaker attractors, by nature, are capable of capturing global speaker characteristics as well as information specific to the particular reference speech sample, since they are trained to extract a particular speaker's speech from a mixed or noisy recording. While there are numerous methods for speech separation , the deep clustering \cite{hershey2016deep} and deep attractor \cite{chen2017deep} based methods are of most relevance to our work. Speaker attractor network \cite{jiang2020speaker}, an extension of the deep attractor network, which employs metric learning among attractors of same speaker in different mixtures to better capture global speaker features along with specific utterance features. 

The key contributions of our work are as follows

\begin{itemize}
    \itemsep0em 
    \item Speaker attractors are extracted from a reference encoder pretrained for speech separation
    \item We adapt the end-to-end TTS method VITS for the purpose of zero shot multi speaker TTS
    \item We show through various experiments that SATTS performs at par with or slightly better than a strong baseline system under clean and challenging recording conditions.
    \item We show, for the first time to the best of our knowledge, the ability to extract speaker embeddings for TTS from mixed reference signals with more than one speakers.
\end{itemize}

\section{Speaker attractor text to speech}

\begin{figure*}[ht]
\setlength\abovecaptionskip{-0.2\baselineskip}
\setlength\belowcaptionskip{-1.3\baselineskip}
\centering
\includegraphics[width=\textwidth]{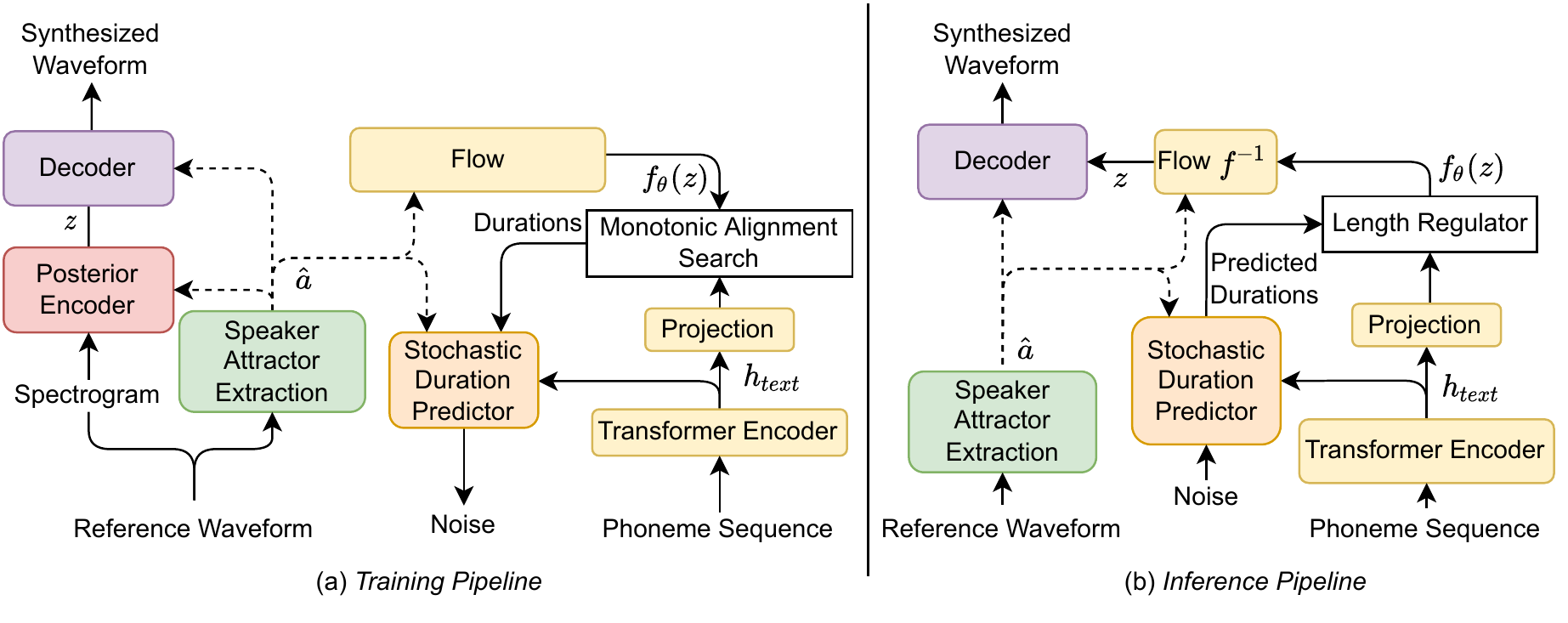}
    \caption{System overview of SATTS during training and inference.}
\label{fig:satts}
\end{figure*}

In this section we describe our proposed speaker attractor text to speech system. The overall steps of SATTS is similar to SV2TTS \cite{jia2018transfer}, in that we first train a model to extract speaker embeddings followed by training of the TTS system with the extracted embeddings. However, in this work, we extract the speaker attractors trained to separate speech from a mixed signal. We utilize these powerful speaker attractors to condition the TTS system. Also unlike SV2TTS, we adapt the end-to-end VITS TTS system, which is a non-autoregressive conditional variational autoencoder. Fig. \ref{fig:satts} shows the system architecture of SATTS. Details about individual components are presented in the following sucsections.

\subsection{End-to-end text to speech}
The TTS backbone in SATTS is adapted from the VITS architecture \cite{kim2021conditional}. We would like to point out that the use of speaker attractors is not limited to just the end-to-end models and can easily be adapted into any existing multi speaker TTS pipeline.


The VITS architecture at its core is a conditional variational autoencoder.
It consists of a posterior encoder, a prior encoder, a stochastic duration predictor and a decoder.

The posterior encoder is a stack of non-causal wavenet residual blocks which takes full scale linear spectrogram as input and produces latent variables as part of a factorized normal distribution, $z$. To enable multi speaker synthesis, the speaker attractor is incorporated as global conditioning to the residual blocks.

The prior encoder consists of a stack of transformer encoder layers as a text encoder which produces the hidden representation $h_{text}$ also parameterized as a factorized normal distribution. The hidden text representation and the latent variables from the posterior encoder are aligned via a hard monotonic alignment matrix at training time. 
A normalizing flow $f_{\theta}$ \cite{rezende2015variational} transforms the hidden text representation $h_{text}$ into a more complex distribution of the posterior in an invertible way, $f^{-1}_{\theta}(f_{\theta}(z))$, by change of variable on top of the factorized normal distribution. The flow is a stack of affine coupling layers with stacks of wavenet residual blocks. Similar to the posterior encoder, the speaker attractor is incorporated as global conditioning to the residual blocks.

The alignment search operation between the prior and posterior distributions is done via Monotonic Alignment Search (MAS)  \cite{kim2020glow}, that searches for the alignment which maximizes the Evidence Lower Bound (ELBO) of data parameterized by a normalizing flow.

In conjunction with the prior encoder and MAS, a stochastic duration predictor (SDP) is trained to estimate the duration of $h_{text}$ for each time step. During inference, since the posterior is not available, the predictions from the stochastic duration predictor are used to regulate the length of the hidden text representation before feeding them into the inverse flow. The stochastic duration predictor is a flow based generative model which is trained via a variational lower bound of the log-likelihood of the phoneme duration, which is provided by MAS. For a more in-depth discussion about training the stochastic duration predictor, please refer to \cite{kim2021conditional}.

The decoder ($G$) architecture is akin to the HiFi-GAN \cite{kong2020hifi} generator which consists of a stack of transposed convolution layers followed by multi receptive field receptors. The decoder takes the latent $z$ and upscales it as per the hop size of the spectrogram operation. To train the decoder efficiently and reduce its memory footprint, random fixed length slices are extracted from $z$. The speaker attractor is linearly transformed and added to the latent variable $z$. 

The proposed SATTS model is trained with the same objective functions as VITS. 
\setlength{\belowdisplayskip}{5pt} \setlength{\belowdisplayshortskip}{5pt}
\setlength{\abovedisplayskip}{5pt} \setlength{\abovedisplayshortskip}{5pt}
\begin{equation}
    L_{tts} = L_{recon} + L_{kl} + L_{dur} + L_{adv}(G) + L_{fm}(G),
\end{equation}
where $L_{recon}$ is the mel-spectrogram reconstruction loss, $L_{kl}$ is the KL-divergence between the prior and posterior distributions following expansion of the prior with MAS, $L_{dur}$ is the negative of the variational lower bound of log-likelihood of the phoneme duration for the stochastic duration predictor and $L_{adv}$ and $L_{fm}$ are the adversarial and discriminator feature matching losses given by a set of multi-period discriminators as in HiFi-GAN. For a more detailed description of the loss terms please refer to \cite{kim2021conditional}.

\subsection{Speaker attractors}

\begin{figure*}[ht]
\setlength\belowcaptionskip{-1.3\baselineskip}
\centering
\includegraphics[width=0.85\linewidth]{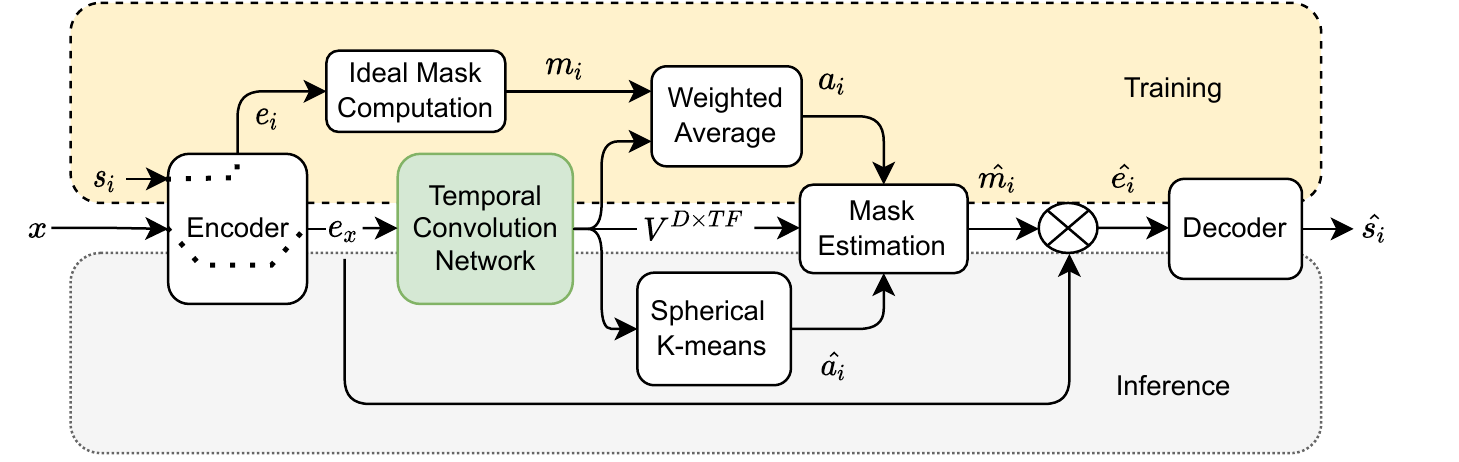}
    \caption{Overall architecture of Speaker Attractor Network}
    \label{fig:sanet}
\end{figure*}



Speaker attractors \cite{jiang2020speaker} are points in high dimensional embedding space which pull time-frequency bins belonging to that speaker towards itself in mixed or corrupted signals. Speaker attractors are trained such that they are able to separate speaker sources from a mixture while also being localized in the global embedding space. This is important, as this enables the speaker attractors to be used in cases where the number of sources in the mixture can be different (more or less) than what was used during training, thus generalizing well to unknown number of sources. This property is also the cornerstone of SATTS, as this allows us to use the speaker attractors for capturing a holistic representation of the target speaker from the reference utterance. 

The training and inference pipelines of the speaker attractor network (SANET) are shown in Fig. \ref{fig:sanet}. It consists of a temporal convolution network, as the separation backbone, encapsulated by an encoder and a decoder. 

While the encoder and decoder can be any function which can extract a time frequency representation of audio waveform, it has been shown that a data driven pre-trained encoder decoder performs best for speech separation \cite{tzinis2020two}. The encoder comprises of a single convolution layer followed by rectified linear unit activation, while the decoder is a single transposed convolution layer without any activation. The time domain waveforms, are processed with overlapping windows of 16 samples with a hop of 8 samples producing a sequence of frames with a $F$ dimensional vector per frame, $e$. This form of time frequency representation is quite useful because the encoding and decoding process does not have to worry about phase reconstruction \cite{takahashi2018phasenet}. The encoder and decoder are first pretrained without the TCN. 

The separation backbone TCN is an adapted version of the TCN in Conv-TasNet \cite{luo2019conv}. It takes the time frequency input embeddings, $e_x$, of the mixture signal $x$, and produces $D$ dimensional vectors for each time frequency ($TF$) bin, $V^{D\times TF}$ 

During training, the time frequency representation ($e_i$) of the $C$ mixing sources, ($s_i$), are used to compute an ideal ratio mask \cite{liutkus2015generalized}, $m_i$. These ideal masks are then used for weighted averaging of $V^{D\times TF}$ to produce the ideal attractors, $a_i$, which lie on the unit sphere in $\mathbb{R}^D$, for each source in the mixture.
\begin{equation}\label{eq:attractor}
\boldsymbol{a}_{i}=\frac{\boldsymbol{V} \cdot\left(\boldsymbol{w} \odot \boldsymbol{m}_{i}\right)}{\left\|\boldsymbol{V} \cdot\left(\boldsymbol{w} \odot \boldsymbol{m}_{i}\right)\right\|_{2}},
\end{equation}
where, $\boldsymbol{w}=\boldsymbol{e}_{x} /\left\|\boldsymbol{e}_{x}\right\|_{1}$ is the weight which ensures that low energy regions (silence) do not affect the attractor formation.

Following this, the cosine distance between the vectors of $V$ are computed to each attractor followed by a $C$ way softmax over the distances to decide the assignment of each time frequency bin to the closest attractor. This operation produces the estimated masks, $\hat{m_i}$ which are then applied to $e_x$ and decoded by the decoder to produce the source estimates, $\hat{s_i}$.

The SANET model is trained with a combination of three objective functions, a reconstruction loss, a contrastive circle loss and a compactness loss,
\begin{equation}
{L}=L_{recon} + L_{circle} + L_{compact}
\end{equation}
The reconstruction loss is the scale invariant signal to distortion ratio between the sources and estimates. The contrastive circle loss drives the attractors to be localized in the global embedding space. And, the compactness loss leads to a compact distribution of embedding vectors of the same speaker. For detailed description of each loss term, please refer to \cite{jiang2020speaker}.

During inference, the attractors ($\hat{a_i}$) are estimated from $V^{D\times TF}$ by means of Sperical K-means clustering \cite{dhillon2001concept}, which uses cosine distance instead of Euclidean distance, which ensures there is no mismatch between training and inference time.

For the purpose of extracting the speaker attractors of the reference waveforms for SATTS training, we set $K=1$ for spherical K-means clustering.

\section{Experiments}

In our experiments, we compare SATTS with a SV2TTS baseline system where we swap out the speaker attractor extraction with a speaker encoder trained for the speaker verification task, similar to \cite{jia2018transfer}. The baseline speaker encoder is prepared with a ResNet backbone and trained with Angular Prototypical \cite{chung2020in} loss function. 

Both the speaker encoder of SV2TTS and SANET are trained on the English \textit{train} subset of the Commonvoice v6.1 dataset \cite{commonvoice:2020}, which is a large scale crowdsourced speech dataset, with 66k speakers and a variety of accents and recording conditions. We resample the speech dataset to 16kHz sampling rate and use a speaker embedding dimension $D=128$ for both the methods.

For training the SANET, we created the mixture by adding two randomly chosen utterances from the dataset and scaling them with random gains $r$ and $(1-r)$ with $0.25\leq r \leq 0.75$



To reduce the training time, similar to \cite{Casanova2021yourtts}, we initialized  the TTS model weights from a single speaker model trained on LJSpeech dataset \cite{ljspeech17} for 1 million steps, followed by multi-speaker training on the \textit{train-clean-100} subset of the LibriTTS dataset \cite{zen2019libritts}, at 22050Hz. To generate the spectrograms for the posterior encoder, a 1024 point short time Fourier transform (STFT) with sliding windows of 1024 samples and 75\% overlap is used. The input to the text encoder is the IPA phonetic transcription of the text. For training the decoder we use segments of 32 frames and 80 mel bands.

  We train each TTS model on 4 Nvidia A100 GPUs with 80GB of memory, with a batch size of 108 per GPU for a total batch size of 432. We used AdamW optimizer with $\beta_1 = 0.8, \beta_2 =
0.99$ and weight decay $\lambda = 0.01$. The learning rate is exponentially reduced by a factor of $0.999875$ with an initial learning rate of $2e^{-4}$. We use mixed precison training \cite{micikevicius2018mixed} and train the models for a further 40k iterations.

We perform inference on a total of 21 unseen speakers, 10 from \textit{test-clean} subset of the LibriTTS and 11 from VCTK dataset \cite{yamagishi2019cstr}. There are 12 female and 9 male speakers.
We randomly draw 55 test sentences from the \textit{test-clean} subset of the LibriTTS dataset, with a constraint of at least 20 words per sentence. 5 utterances were synthesized per speaker. 
As ground truth, we randomly select 5 audios for each of the test speakers. We set the noise scaling parameters of both the prior encoder and the stochastic duration predictor (SDP) to 0.333 for all experiments, except that in Sec.\ref{exp:mix}, which uses 1.333 for the SDP. Please refer to the `demo' folder in the attached multimedia files to listen to samples.

We used crowd sourced subjective tests to evaluate the naturalness in terms of mean opinion score (MOS) \cite{rec1996p} on a scale of 1-5 with intervals of 0.5. Similar to \cite{jia2018transfer}, we also evaluate the speaker similarity in terms of MOS (sim-MOS). Each sample received one vote and all evaluations done independently without directly comparing any of the methods.

\subsection{Reference with clean recording conditions}
Table \ref{tab:clean} shows the performance of SATTS in comparison with SV2TTS under clean recording conditions. All reference samples were taken from the dataset and resampled to 16kHz for extraction of the speaker embeddings and attractors followed by TTS inference. We observe that SV2TTS has a slight advantage over SATTS in terms of sim-MOS especially for the VCTK dataset, which consists of the most variance in terms of accents. The naturalness MOS is at par or slightly better for SATTS in samples from both the datasets.
\vspace{-1mm}
{\renewcommand{\arraystretch}{1.2}
\begin{table}[H]
\setlength{\tabcolsep}{2.5pt}
\centering
\caption{Comparison of MOS and sim-MOS between SV2TTS and SATTS outputs for clean reference signals.}
\begin{tabular}{l|c|c|c|c} 
\hline
       & \multicolumn{2}{c|}{MOS}                    & \multicolumn{2}{c}{sim-MOS}                  \\ 
\cline{2-5}
       & LibriTTS             & VCTK                 & LibriTTS             & VCTK                  \\ 
\hline
GT     & 4.13 ± 0.22          & 4.09 ± 0.15          & 3.88 ± 0.29          & 3.83 ± 0.27           \\ 
\hline
SV2TTS & 3.82 ± 0.27          & \textbf{3.83 ± 0.21} & \textbf{3.68 ± 0.34} & \textbf{3.68 ± 0.28}  \\ 
\hline
SATTS  & \textbf{3.95 ± 0.21} & 3.79 ± 0.25          & 3.66 ± 0.27          & 3.45 ± 0.32           \\
\hline
\end{tabular}
\label{tab:clean}
\vspace{-1mm}
\end{table}
}
\subsection{Reference with varying recording conditions}
\vspace{-2mm}
{\renewcommand{\arraystretch}{1.2}
\begin{table}[H]
\setlength{\tabcolsep}{3pt}
\centering
\caption{Comparison of MOS and sim-MOS between SV2TTS and SATTS outputs when the reference signal is convolved with a random RIR, simulating different recording conditions.}
\begin{tabular}{l|c|c|c|c} 
\hline
       & \multicolumn{2}{c|}{MOS}                   & \multicolumn{2}{c}{sim-MOS}                  \\ 
\cline{2-5}
       & LibriTTS            & VCTK                 & LibriTTS             & VCTK                  \\ 
\hline
GT-RIR & 4.16 ± 0.22         & 3.91 ± 0.22          & 3.54 ± 0.36          & 3.50 ± 0.31            \\ 
\hline
SV2TTS & 3.91 ± 0.17         & 3.96 ± 0.20           & 3.38 ± 0.36          & 3.50 ± 0.34            \\ 
\hline
SATTS  & \textbf{3.96 ± 0.20} & \textbf{4.06 ± 0.21} & \textbf{3.47 ± 0.34} & \textbf{3.62 ± 0.31}  \\
\hline
\end{tabular}
\label{tab:rir}
\vspace{-1mm}
\end{table}
}
To evaluate the synthesis performance of the proposed method under different recording and reverberation conditions, for each of the test reference samples and ground truths, we randomly sample one simulated room impulse responses from the \textit{simulated\_rirs} subset of the Room Impulse Response and Noise dataset \cite{ko2017study}. We perform the exact same evaluation as with the cleanly recorded samples described above. All samples are compared with the clean version of the reference for the subjective evaluation. Table \ref{tab:rir} show the performance comparison under this condition. We can see that SATTS performs slightly better than SV2TTS in both the datasets. Though, it must be noted that SV2TTS also works quite well as the verification training is done on the Commonvoice dataset which has a variety of recording conditions.

\subsection{Reference with overlapping speakers} \label{exp:mix}
We evaluate the performance of SATTS in zero shot speaker adaptation when the reference sample consists of a mix of more than one speaker. We added a distractor speech sample from a different unseen speaker from the LibriTTS dataset to all the test reference signals, and extracted the target speaker's attractor by setting $K=2$ for the spherical K-means clustering. Since extraction of speaker embedding from mixed speech is not possible for SV2TTS, we only evaluate SATTS. Table \ref{tab:mix} shows the performance comparison of the clean, RIR and mix conditions. We can clearly see that SATTS works in these challenging and different settings without compromising on the naturalness and speaker similarity. While, it might seem odd that the clean reference is not the best performing one, this is expected, since the mixed condition matches the training of SANET better than the clean condition.
\vspace{-1mm}
{\renewcommand{\arraystretch}{1.2}
\begin{table}[H]
\setlength{\tabcolsep}{4pt}
\centering
\caption{Comparison of MOS and sim-MOS for SATTS outputs based on the conditions of reference signal, clean speech, convolved with random RIR and mixed with another speech.}
\begin{tabular}{l|c|c|c|c} 
\hline
       & \multicolumn{2}{c|}{MOS}                                                            & \multicolumn{2}{c}{sim-MOS}                                                  \\ 
\cline{2-5}
       & LibriTTS                                 & VCTK                                     & LibriTTS                         & VCTK                                      \\ 
\hline
Clean & 3.95 ± 0.21                              & 3.79 ± 0.25                              & \textbf{3.66 ± 0.27}             & 3.45 ± 0.32                               \\ 
\hline
RIR    & \multicolumn{1}{l|}{\textbf{3.96 ± 0.20}} & \multicolumn{1}{l|}{4.06 ± 0.21}         & \multicolumn{1}{l|}{3.47 ± 0.34} & \multicolumn{1}{l}{\textbf{3.62 ± 0.31}}  \\ 
\hline
Mix    & \multicolumn{1}{l|}{3.84 ± 0.26}         & \multicolumn{1}{l|}{\textbf{4.09 ± 0.20}} & \multicolumn{1}{l|}{3.60 ± 0.32}  & \multicolumn{1}{l}{3.40 ± 0.31}            \\
\hline
\end{tabular}
\label{tab:mix}
\vspace{-1mm}
\end{table}
}

\subsection{Speech recognition results}
\vspace{-1mm}
{\renewcommand{\arraystretch}{1.2}
\begin{table}[H]
\setlength{\tabcolsep}{2pt}
\centering
\caption{Comparison word error rate (WER) based on a off-the-shelf ASR model from SpeechBrain (Lower is Better)}
\begin{tabular}{c|c|c|c|c} 
\hline
    & SV2TTS-clean & SV2TTS-rir & SATTS-clean   & SATTS-rir  \\ 
\hline
WER & 8.65         & 8.74       & \textbf{6.30} & 6.34       \\
\hline
\end{tabular}
\label{tab:asr}
\vspace{-1mm}
\end{table}
}
We also compared the intelligibility in terms of automatic speech recognition (ASR) performance using an off-shelf ASR model from SpeechBrain \cite{speechbrain}. Table \ref{tab:asr} shows that SATTS achieves much better word error rates (WER) compared to the baseline SV2TTS method for all evaluations samples from both LibriTTS and VCTK dataset demonstrating the superiority of speaker attractors for TTS.

\section{Conclusion}
In this work, we propose speaker attractor text to speech which utilizes speaker attractors trained for speech separation and transfer the learning to train an end-to-end text to speech synthesis system. Through subjective evaluations, we show the robustness of SATTS over various challenging conditions for the reference signal. To the best of our knowledge, this is the first work to utilize transfer learning from speech separation to text to speech synthesis, and the ability to extract target speaker's attractor from a signal with more than one speakers speaking simultaneously could be quite useful in real world applications.



\section{Acknowledgements}

This work was partially supported by JST AIP Acceleration Research JPMJCR20U3, Moonshot R\&D Grant Number JPMJPS2011, JSPS KAKENHI Grant Number JP19H01115, and JP20H05556 and Basic Research Grant (Super AI) of Institute for AI and Beyond of the University of Tokyo.

\bibliographystyle{IEEEtran}

\bibliography{template}



\end{document}